\documentclass{appolb}
\usepackage{epsfig}
\usepackage{graphicx}
\usepackage{wrapfig}

\newcommand{\snn} {\mbox{$\sqrt{s_{NN}}$}}

\def\Journal#1#2#3#4{#1 {\bf #2}, #3 (#4)}
\def\JournalAPPEAR#1#2{#1 {\bf #2}}

\def\JPG{J.~Phys.~G}

\def\NPA{Nucl. Phys.~A}

\def\PLB{Phys. Lett.~B}
\def\PRL{Phys. Rev. Lett.}
\def\PRC{Phys. Rev.~C}

\newcommand{\axi}{$\overline{\Xi}^+$}
\newcommand{\xim}{$\Xi^-$}
\newcommand{\alam}{$\overline{\Lambda}$}
\newcommand{\lam}{$\Lambda$}
\newcommand{\ks}{$\mathrm{K}^{0}_{S}$}
\newcommand{\omm}{$\Omega^-$}
\newcommand{\aom}{$\overline{\Omega}^+$}
\newcommand{\pt}{$p_{T}$}

\begin{document}
\pagestyle{plain}
\newcount\eLiNe\eLiNe=\inputlineno\advance\eLiNe by -1
\title{Production and Flow of Identified Hadrons at RHIC}
\author{Julia Velkovska
\address{Vanderbilt University, Nashville, TN 37235, USA}}
\maketitle

\vspace{-0.5cm}

\begin{abstract}
We review the production and flow of identified hadrons at RHIC with a
main emphasis on the intermediate transverse momentum region( $p_T
\approx$ 2--5 GeV/$c$). The goal is to unravel the dynamics of baryon
production and resolve the anomalously large baryon yields and
elliptic flow observed in the experiments.
\end{abstract}


\vspace{-0.5cm}

\section{Introduction}
This paper explores the relativistic heavy ion collisions at RHIC and
the medium produced in these collisions using hadronic
observables. Being most abundantly produced, hadrons define the bulk
medium behavior, which is governed by soft, non-perturbative particle
production.  Analysis of identified hadron spectra and yield ratios
allows determination of the kinetic and chemical properties of the
system. Hydrodynamics models~\cite{hydro} have been successful
in reproducing identified hadron spectral shapes and their
characteristic mass dependence at low \pt as well as 
the azimuthal anisotropy of particle emission. Notably, in order to match the data the models 
require rapid equilibration of the produced matter and
a QGP equation of state.  The particle abundances also point to an equilibrated 
system and are well described by statistical thermal models~\cite{thermal}.
The chemical freeze-out at $T_{ch} \approx 170$ MeV is suggestive, as it is at the phase
boundary of the transition between hadron gas and QGP, as predicted by
lattice QCD calculations~\cite{Lqcd}.

Above $p_{T}\approx 2$ GeV/$c$, hard-scattering processes become increasingly
important. After the hard-scattering, a colored object (the
hard-scattered quark or gluon) traverses the medium produced in the
collision and interacts strongly. As a result, it loses
energy via induced gluon radiation.  This phenomenon, known as
jet-quenching, manifests itself as suppression in the yields of
high-$p_{T}$ hadrons, when compared to the production in pp collisions and 
weakening of the back-to-back angular correlations between the jet
fragments.  The yield suppression is measured in terms of the nuclear
modification factor $R_{AA} = Yield_{AA}/N_{coll}/Yield_{pp}$, where
the number of binary nucleon-nucleon collisions, $N_{coll}$, is
introduced to account for the nuclear geometry. In this paper, we use
the ratio $R_{CP}$, which is obtained from the $N_{coll}$ scaled
central to peripheral spectra and carries similar information. Jet
quenching was discovered at RHIC both in suppressed hadron production
at high-$p_{T}$ ~\cite{ppg003} ($R_{AA}<1$) and in vanishing
back-to-back jet correlations~\cite{starb2b}.

Another discovery, unpredicted by theory, is a large enhancement in
the production of baryons and anti-baryons at intermediate $p_{T}
\approx$ 2--5 GeV/$c$~\cite{ppg006,ppg015},compared to expectations
from jet fragmentation.  This is in contrast to the suppression of $\pi^{0}$~\cite{ppg014}.
In central $Au+Au$ collisions the
ratio $\overline{p}/\pi$ is of the order 1 - a factor of 3  above
the ratio measured in peripheral reactions or in $pp$ collisions. In
this region of $p_{T}$ fragmentation dominates the particle production
in $pp$ collisions. It is expected that fragmentation is independent
of the colliding system - hence the large baryon fraction observed at
RHIC comes as a surprise.  At RHIC,  the medium
influences the dynamics of hadronization resulting in enhanced baryon
production but the exact mechanism is not yet completely understood.  
This paper reviews the latest experimental results relevant to this subject.
%
\section{Radial flow at intermediate $p_{T}$.} 
 The most common conjecture that is invoked to explain the large
 $\overline{p}/\pi$ ratios observed by PHENIX ~\cite{ppg015} is the
 strong radial flow that boosts the momentum spectra of heavier
 particles to high $p_{T}$.  In this scenario, the soft 
\begin{figure}[h]
\begin{center}
\includegraphics[width=0.8\linewidth]{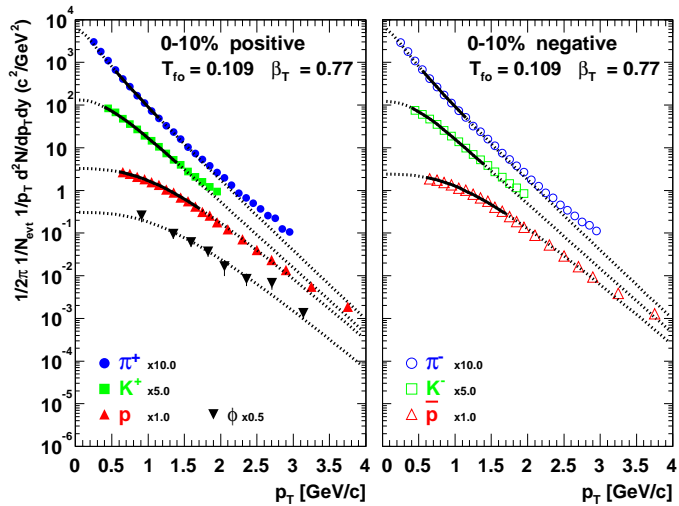}
\caption{ Transverse momentum spectra of $\pi^{\pm}, K^{\pm}$, and
$p,\overline p$ and hydrodynamical fit results for 0--10\% central
Au+Au collisions at $\snn = $ 200 GeV~\cite{ppg016}. The \pt ranges for 
the fit are indicated by the solid lines, while the 
dashed lines show the extrapolated predictions for each particle species. The
$\phi$-meson spectrum, not included in the fit, is compared to the
model's prediction.}
\label{fig:radflow}
\end{center}
\end{figure}   
processes
 dominate the production of (anti)protons at 2--4.5 GeV/$c$, while the
 pions are primarily produced by fragmentation of hard-scattered
 partons.  In Fig.~\ref{fig:radflow} we compare the 10\% central
 spectra of $\pi^{\pm}, K^{\pm}$,$ p,$ and $\overline p$ to a
 hydrodynamics  model~\cite{Schnedermann93} that has been
 fitted to the data. The free parameters in the model are the kinetic
 freeze-out temperature $T_{fo}$, the transverse flow velocity
 $\beta_{T}$ and the absolute normalization. The line drawn through
 the $\phi$-meson spectrum is the model's prediction obtained after
 fitting all other particle species. We see that: 1) Hydrodynamics
gives a good description of the $p$ and $\overline p$ spectral shapes
 up to $\approx 3$ GeV/$c$, and 2) the $\phi$-meson spectrum can be
 described by the same parameter set as the protons.For lighter
 particles, the deviation from hydrodynamics happens at lower $p_{T}$.
These results may lead to the conclusion that the enhanced
 $\overline{p}/\pi$ ratio is a mass effect and the intermediate \pt 
 (anti)protons are primarily produced in soft processes. 
We now examine the scaling of the yields in different centrality
classes. We expect that for soft production the yields will scale as
the number of nucleons participating in the collision, while for hard
processes the scaling is with $N_{coll}$.  
In Fig.  ~\ref{fig:scaling} the \pt distributions for
(anti)protons and $\phi$ are scaled down each by their respective
$N_{coll}$.  To isolate mass effects from baryon/meson effects we
compare a heavy meson to the protons. The result is rather
surprising.  At intermediate \pt the $p+\overline{p}$ yields scale with
$N_{coll}$ as expected for hard processes.
\begin{figure}
\begin{center}
\includegraphics[width=0.8\textwidth]{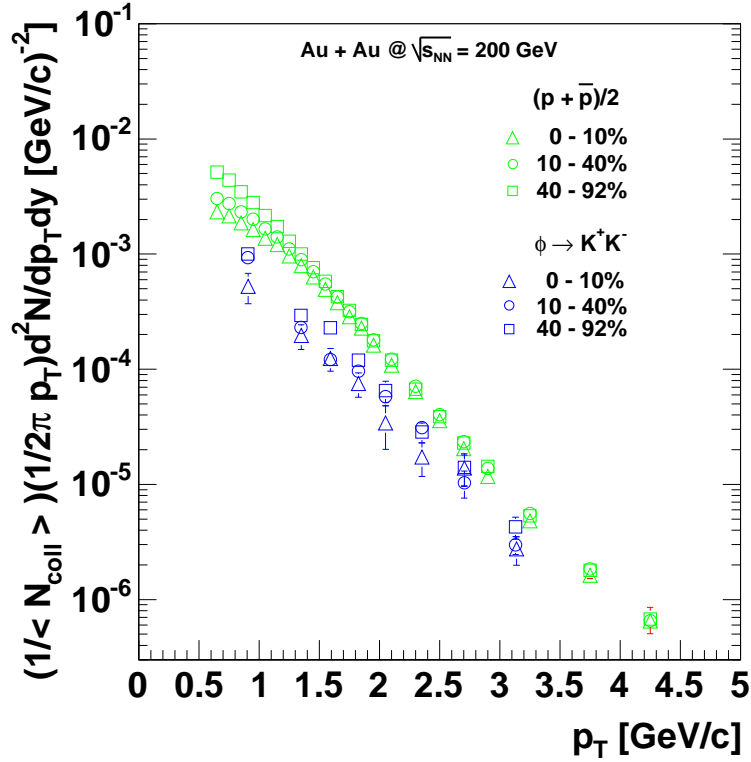}
\caption{ $N_{coll}$-scaled transverse momentum spectra of
$(p+\overline {p})/2$ and $\phi$-mesons for three different centrality
classes. High-$p_{T}$ baryon yields scale with
$N_{coll}$, while no scaling is observed for $\phi$-mesons~\cite{ppg016}.}
\label{fig:scaling}
\end{center}
\end{figure}
 The $\phi$ yields do
not scale. Although the shape of the $p+\overline{p}$ and $\phi$
spectra is the same and is well reproduced by hydrodynamics, the
absolute yields for $\phi$ grow slower with centrality. When the
central and peripheral yields are used to evaluate the nuclear
modification factor (Fig.  ~\ref{fig:phircp}), the
(anti)protons show no suppression ($R_{CP} \approx 1$) , while the $\phi$ are
suppressed similar to $\pi^{0}$. This result rules out the radial flow
(and the mass) as the sole factor that is responsible for the baryon
enhancement. The similarity in the centrality dependence of $\phi$ and
$\pi$ production suggests an effect related to the number of
constituent quarks rather than the mass. The STAR experiment also
observed a clear baryon/meson distinction in $R_{CP}$ of $K^{*}$,
$K^{0}_{s}$, $\Lambda$, and $\Xi$~\cite{starLambda}.
\begin{figure}[b]
\begin{center}
\includegraphics[width=0.9\textwidth]{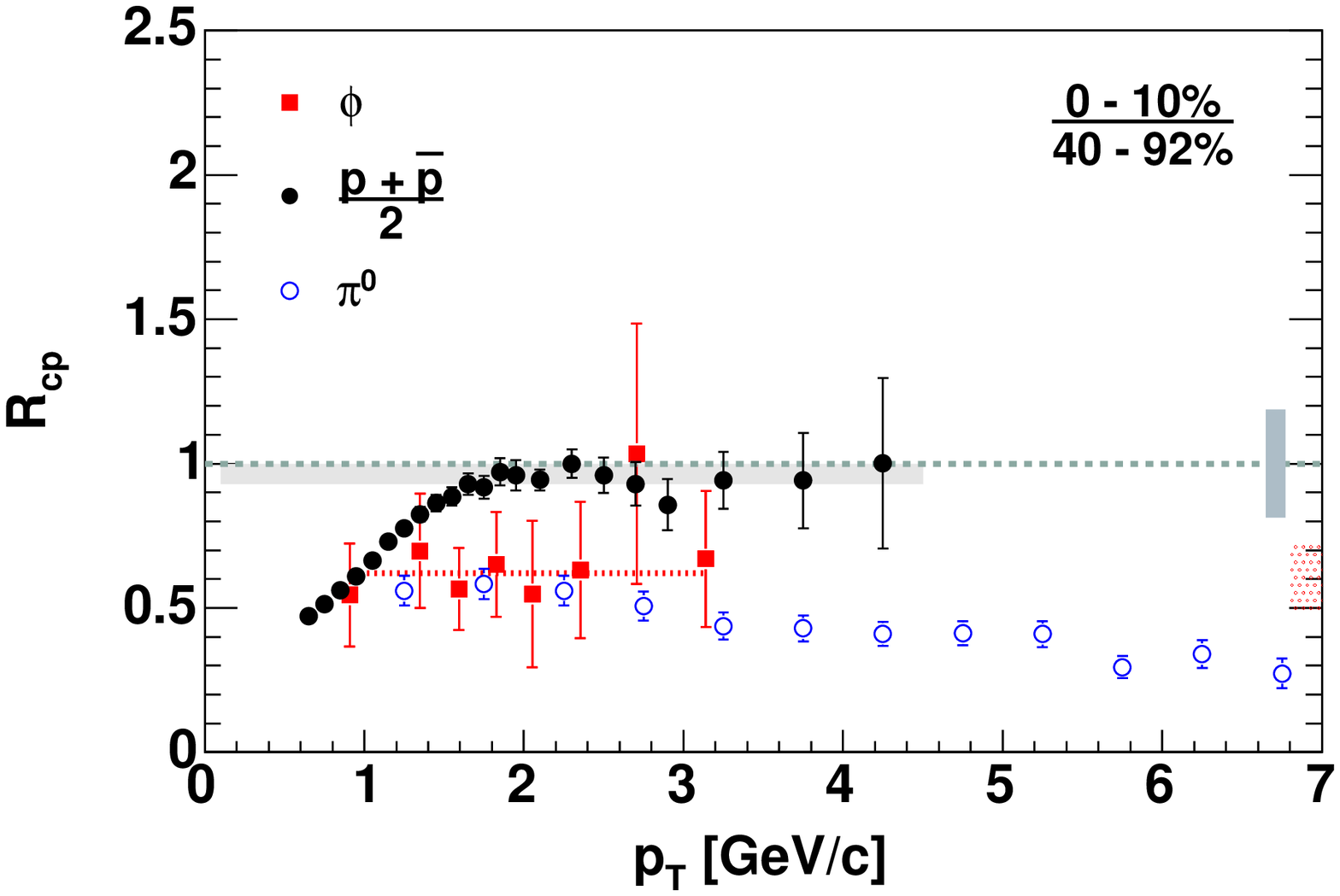}
\caption{The nuclear modification factor, $R_{CP}$, of  $\phi$, $p+\overline{p}$ and $\pi^{0}$ 
 measured by PHENIX in  Au+Au collisions at $\snn = $ 200 GeV~\cite{qmproc,ppg016}.}
\label{fig:phircp}
\end{center}
\end{figure}
\section {Recombination and empirical scaling of elliptic flow.} 
\begin{figure}
\begin{minipage}[t]{0.48\textwidth}
\begin{center}

\includegraphics[width=0.99\textwidth]{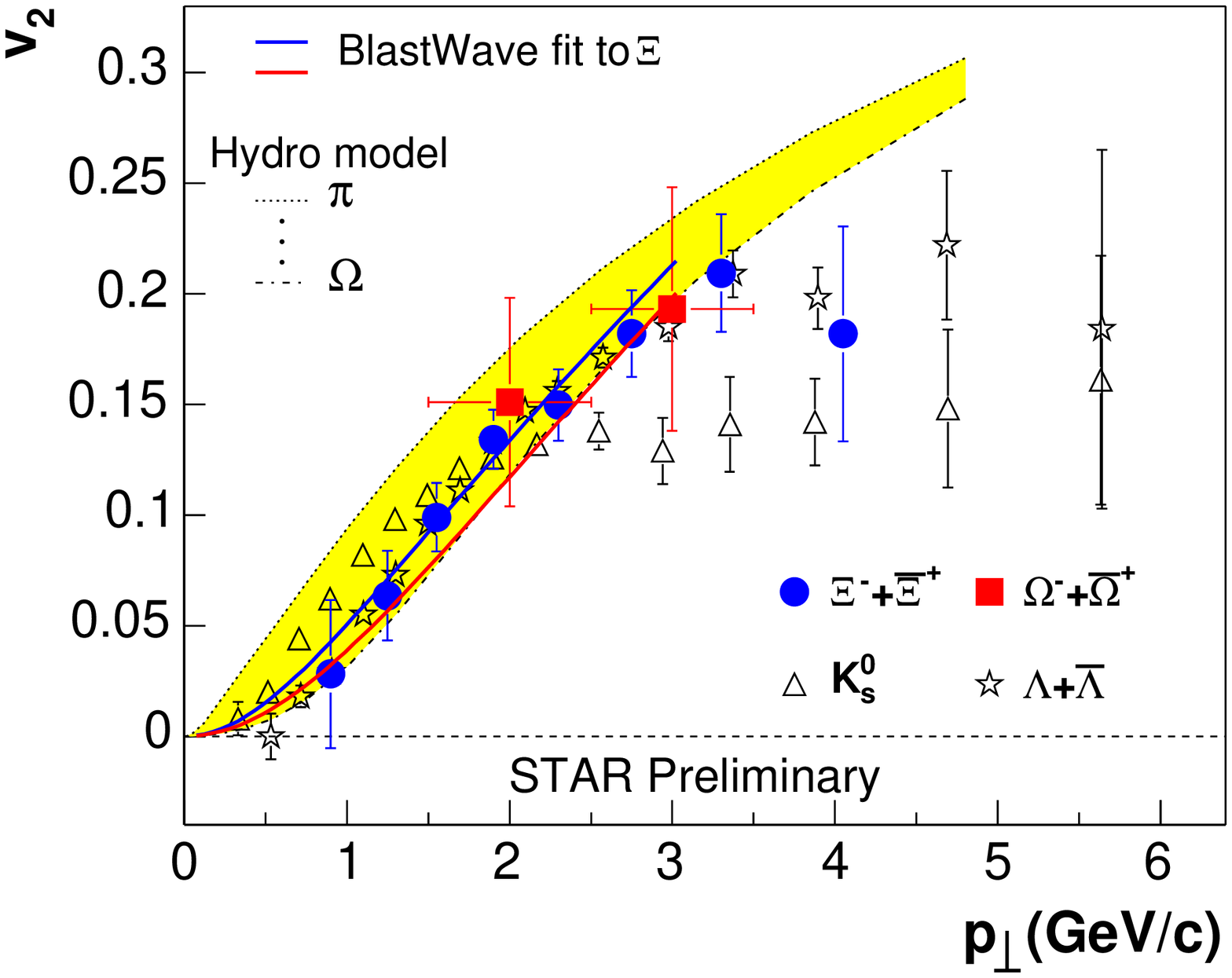}
\caption{$v_2(p_{T})$ for \xim+\axi , \omm+\aom , \ks ~and \lam+\alam
for minimum bias $Au+Au$ collisions~\cite{Castillo}.  The curves show
the results from hydrodynamics calculations.}
\label{Fig:V2PtXiOm}
\end{center}
\end{minipage}\hfill
\begin{minipage}[t]{0.48\textwidth}
\begin{center}
\includegraphics[width=0.99\textwidth]{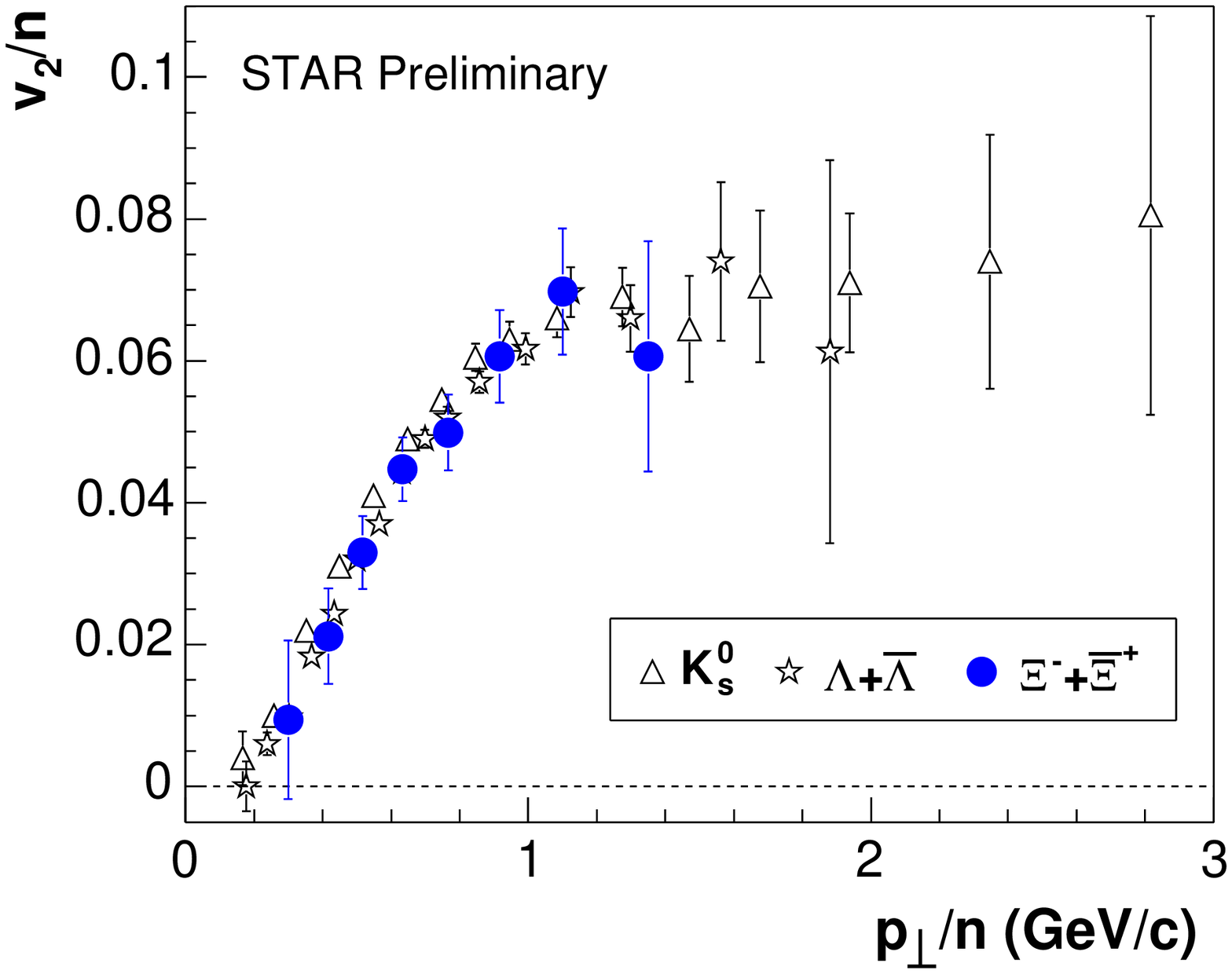}
\caption{$v_{2}/n$ as a function of $p_{T}/n$ for \ks,
\lam+\alam~\cite{starLambda} and \xim+\axi, where $n$ is the number
of constituent quarks for each particle.  (Figure taken
from~\cite{Castillo}.)}
\label{Fig:QuarkCoal}
\end{center}
\end{minipage}\hfill
\end{figure}

Recently, several quark recombination
models~\cite{recoDuke,recoOregon,recoTAMU} have been proposed to resolve
the RHIC baryon puzzle.  In the dense medium produced in central $Au+Au$ collisions,
recombination of quarks becomes a natural hadronization mechanism.  When
produced from the same underlying thermal quark distribution, baryons get
pushed to higher \pt than mesons due to the addition of quark momenta.  At
intermediate \pt recombination wins over fragmentation for baryons, while
mesons are still dominated by fragmentation.  After fitting the inclusive
 hadron spectra to extract the thermal component, the models
are able to reproduce a large amount of data on identified particle
spectra, particle ratios and nuclear modification factors.  The most
spectacular success of the recombination models comes from the comparison
with the data on elliptic flow. At low-\pt hydrodynamics describes both
the magnitude and the mass dependence of $v_{2}$.  However, at \pt$>2$
GeV/$c$ the mass ordering of $v_{2}$ changes,namely : $v_{2}(p) >
v_{2}(\pi)$~\cite{ppg022} and $v_{2}(\Lambda) >
v_{2}(K_{s})$~\cite{starLambda}.  In addition, the size of the signal is
too big to be explained by asymmetric jet absorption~\cite{Molnar03}.The
recombination models solve the problem by assigning the elliptic flow
signal to the quarks, instead of the hadrons.  Then the baryon/meson split
in $v_{2}$ is naturally explained.  It has been demonstrated empirically,
that the flow per quark is universal. Recent results from the STAR~\cite{Castillo}
experiment that include the measurement of multi-strange baryons are shown
in Fig.~\ref{Fig:V2PtXiOm}. A clear baryon/meson difference is observed in
the data at \pt $>2 $GeV/$c$. The results from a typical hydrodynamic model
calculations~\cite{Pasi01} are shown with a band.  After re-scaling of
both axes in Fig.~\ref{Fig:V2PtXiOm} to represent the quark flow, the data
falls on a universal curve as demonstrated in Fig.~\ref{Fig:QuarkCoal}.

\section{Jet correlations with leading baryons or mesons}

The recombination models resolve most of the baryon/meson effects
observed in the data.  However, from spectra, particle
ratios and elliptic flow it is difficult to infer whether the
recombining quarks come from the thermal bath (soft processes) or from
hard-scattering. To unravel the nature of the baryon enhancement and
to test the recombination approach, the PHENIX experiment examined the
two-particle angular correlations with identified meson or baryon
trigger particle~\cite{ppg033}.  The momentum of the triggers was
chosen to be in the range of \pt where baryon/meson differences are
observed ($2.5 < p_T < 4 $ GeV/$c$). For both types of trigger, clear
jet-like angular correlations were observed both on the same side of
the trigger and at $180^{0}$. This result shows that both mesons and
baryons have a significant hard-scattering component at intermediate
\pt, although the $\overline{p}/\pi$ ratios are dramatically different
from fragmentation in the vacuum.  In order to quantify the
similarity between baryon and meson triggered correlations, the yield
of associated particles is integrated at the near-side and the away
side peaks. The results for different centrality classes and colliding
systems are shown in Fig.~6.
There is an increase in partner yields in mid-central Au+Au compared
to the d+Au and p+p collisions. In Au+Au collisions, the near side
yield per {\it meson} trigger remains constant as a function of
centrality, whereas the near-side yield per {\it baryon} trigger
decreases in the most central collisions as expected if a fraction of
the baryons were produced by soft processes such as recombination of
thermal quarks.  The dashed line in Fig.~6
represents an upper limit to the centrality dependence of the jet
partner yield from thermal recombination.

\begin{figure}
\begin{center}
\includegraphics[width=0.99\linewidth]{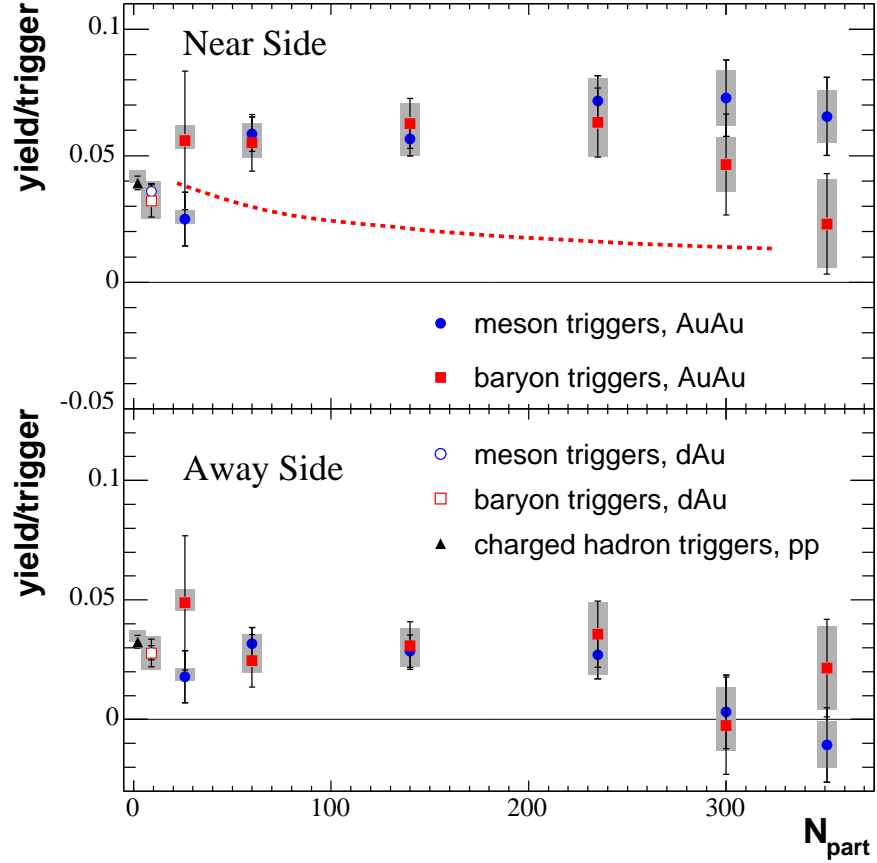}
\caption{Yield per trigger for associated charged hadrons between
$1.7 < p_T < 2.5$ GeV/$c$ for the near- (top) and away- (bottom)  side
jets~\cite{ppg033}. The dashed line (top) represents an upper  limit
of the centrality dependence of the near-side partner yield  from
thermal recombination.}
\end{center}
\label{fig:jet_yield}
\end{figure}   
 The data clearly disagree
with both the centrality dependence and the absolute yields of this
estimation, indicating that the baryon excess has the same jet-like
origin as the mesons, except perhaps in the highest centrality bin.
The bottom panel of Fig. ~6 
shows the conditional
yield of partners on the away side.  It drops equally for both trigger
baryons and mesons going from p+p and d+Au to central Au+Au, in
agreement with the observed disappearance \cite{starb2b} and/or
broadening of the dijet azimuthal correlations.  It further supports
the conclusion that the baryons originate from the same jet-like
mechanism as mesons.The description of the data in the pQCD framework
would require an in-medium modification of the jet fragmentation
functions. For recombination models, the experimental results imply
that shower and thermal partons have to be treated on an equal
basis~\cite{recoOregon}.


\section{Summary}
We reviewed the results on hadron production and flow in relativistic
heavy ion collisions at RHIC.  The production mechanisms at low-\pt and
high-\pt are relatively well understood in terms of soft and hard
processes, respectively.  The intermediate \pt region ($2 < p_{T} < 5
$GeV/$c$) is marked by a number of puzzling experimental observations
and most notably, by the baryon excess over the expectation from vacuum
fragmentation functions.  By comparing spectra and centrality scaling
of (anti)protons and $\phi$-mesons, we established that the excess of
anti-protons with respect to pions is not due to the larger mass of
the anti-proton, but is related to the number of constituent
quarks. The recombination models get a beautiful confirmation in the
empirical scaling relation of the elliptic flow results.
Jet-correlations with trigger baryons or mesons show a similar
hard-scattering component in both.  This observation is also in line
with the $N_coll$ scaling observed in the yields of protons and
anti-protons. However, it implies that protons originate from hadrons
that experience little or no energy loss, while pions come from
partons that have suffered large energy loss. This result is conceptually 
difficult, unless baryons and mesons have a very
different formation time and thus - the original partons have
different time to interact with the medium. Recombination models which
combine hard-scattered partons with thermal ones give the most likely
explanation of the experimental results as a whole. The baryon excess is clearly 
an effect of the medium produced in $Au+Au$ collisions and maybe, thorough 
the comparison with recombination models,  gives evidence for its partonic nature.


\end{document}